# Filtering in CMB data analysis with application to ACT DR4 and *Planck* observations

Erik Rosenberg,[1,2,3,*] Steven Gratton,[4,2,3] and Anthony Challinor[2,3,4]

[1] *Jodrell Bank Centre for Astrophysics, University of Manchester, Oxford Road, Manchester, M13 9PL, United Kingdom*

[2] *Institute of Astronomy, Madingley Road, Cambridge, CB3 0HA, United Kingdom*

[3] *Kavli Institute for Cosmology, Cambridge, Madingley Road, Cambridge, CB3 0HA, United Kingdom*

[4] *DAMTP, Centre for Mathematical Sciences, Wilberforce Road, Cambridge, CB3 0WA, United Kingdom*

Motivated by observed discrepancies between the Atacama Cosmology Telescope Data Release 4 (ACT DR4) and *Planck* 2018 cosmic microwave background (CMB) anisotropy power spectra, particularly in the cross-correlation of temperature and E-mode polarization, we investigate challenges that may be encountered in the comparison of satellite and ground-based CMB data. In particular, we focus on the effects of Fourier-space filtering and masking involving bright point sources. We show that the filtering operation generates bright cross-shaped artifacts in the map, which stretch far outside typical point-source masks. If not corrected, these artifacts can add bias or additional variance to cross-spectra, skewing results. However we find that the effect of this systematic is not large enough to explain the ACT-*Planck* differences presented with ACT DR4.

## I. INTRODUCTION

With the end of operations of the *Planck* satellite [1–3] in 2013 and its final data release in 2020 [4], it is clear that the next decade of cosmic microwave background (CMB) investigation will be conducted from the ground. Ongoing experiments like the Atacama Cosmology Telescope (ACT) [5–9] and the South Pole Telescope (SPT) [10–14]) have lower polarization noise and operate at higher resolution than *Planck*. These, and the forthcoming ground-based experiment Simons Observatory (SO) [15] and the planned next-generation CMB-S4 [16], will significantly improve measurements of the primary CMB polarization as well as the temperature damping tail and small-scale secondary effects such as CMB lensing and the Sunyaev-Zel'dovich effects. However, all of these ground-based experiments are limited in their ability to measure large scales due to atmospheric noise. The atmosphere acts a source of strongly scale-dependent noise, with the noise power spectrum $N_\ell$ rapidly increasing with multipole $\ell$ as $N_\ell \propto \ell^{-3}$ below a given "knee frequency"; in ACT intensity, for example, $\ell_{\text{knee}} \sim 2000$ and 3000 at 98 and 150 GHz, respectively [17]. This noise is worse for higher frequencies, and for intensity measurements relative to polarization. This motivates in part the scale cuts implemented by ACT in Louis *et al.* [8] [$\ell > 500$ (TT), $\ell > 350$ (TE/EE)] and later in Choi *et al.* [18] [$\ell > 600$ (TT), $\ell > 350$ (TE/EE)]. To avoid losing information from the first two CMB peaks while remaining independent of *Planck*, ACT Data Release 4 (DR4) was therefore combined with *WMAP* data [19] at large and intermediate ($2 < \ell < 1200$ in TT and $24 < \ell < 800$ in TE) scales for the baseline likelihood analysis in Aiola *et al.* [20]. Similarly, baseline SO forecasts include *Planck* on large scales and areas of the sky not observed by SO. Higher *Planck* frequencies not accessible from the ground, such as 353 GHz, are also expected to be useful for constraining foreground properties like the spatial variation of the spectral index of Galactic dust [15]. We anticipate that space-based measurements from *Planck* and *WMAP*, and in the future *LiteBIRD* [21,22], will continue to be used in combination with ground-based data for the foreseeable future. It is therefore necessary to study in detail potential problems that may arise in combining these substantially different datasets. In this paper we investigate challenges in the combination of *Planck* and ACT data, drawing conclusions that are more widely applicable to other experiments as well.

The principal artifacts studied here are created by strong point sources subject to a Fourier-space filter applied to remove contamination from the ground. This effect is relatively well known, mentioned in passing and accounted for in recent ACT and SPT papers, though not in all cases. One of our aims in this article is to highlight this effect as an example case of the care that should be used when applying such filters.

[*] Contact author: erik.rosenberg@manchester.ac.uk

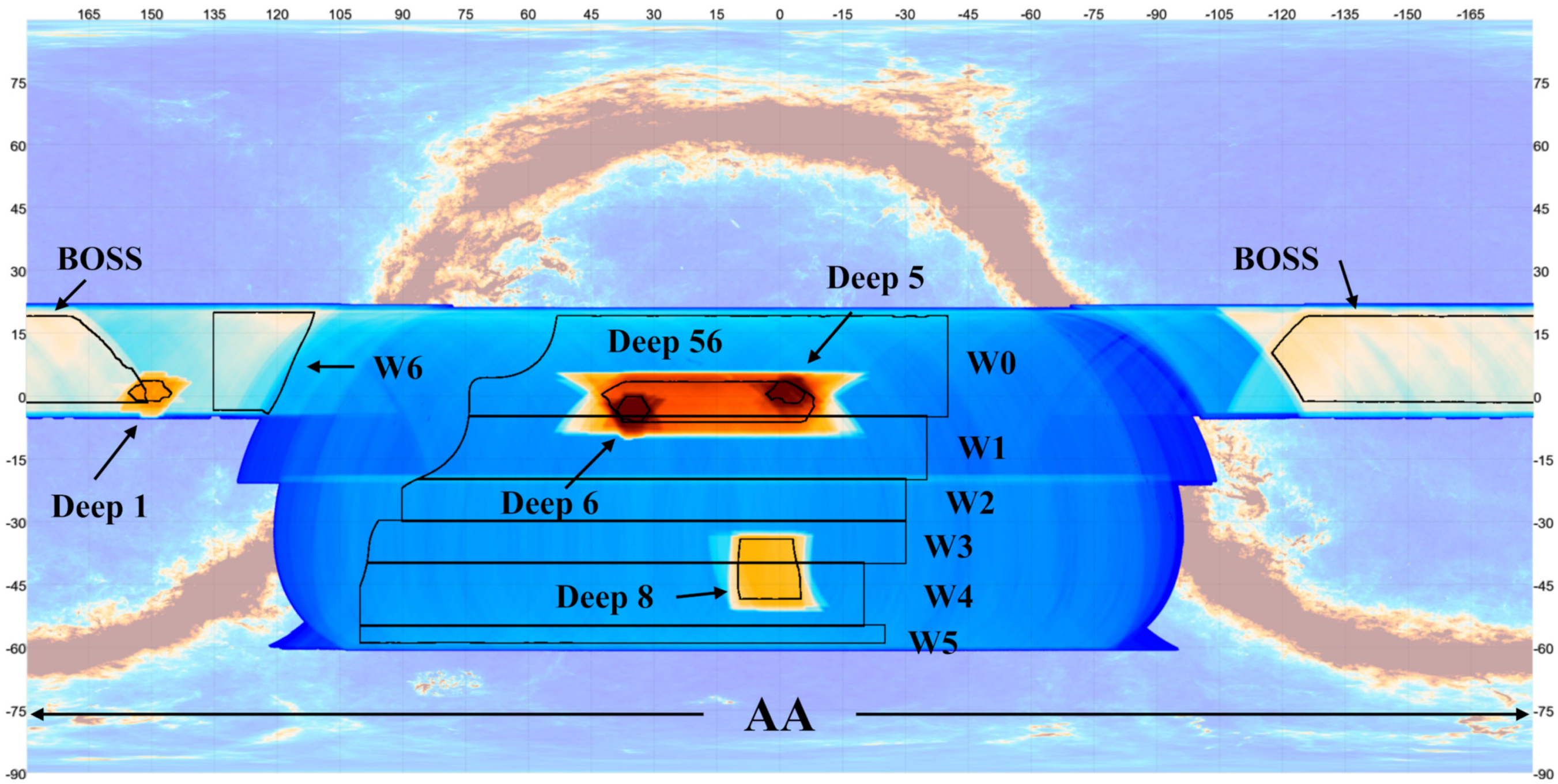

FIG. 1. ACT DR4 sky coverage in equatorial coordinates. The right ascension (horizontal) and declination (vertical) are given in degrees. The color scale shows map depth and the background is *Planck* 353 GHz intensity. In this work we focus on the BOSS and Deep 56 regions. Figure from Choi *et al.* [18] reproduced with permission.

We present the particular context of this study in Sec. II with an introduction to ACT DR4. In Sec. III we focus in particular on the filtering step of that analysis. In this section we also begin to explore the interaction of this filtering procedure with bright point sources and derive analytic results for the resulting artifacts and their impact on the power spectrum. We see in Sec. IV the impact this effect has on power spectra calculated from real data. Finally, some strategies for mitigation are discussed in Sec. V, and we conclude in Sec. VI.

## II. ACT DR4

In this section we place ACT DR4 into context and review the aspects of the DR4 maps and power spectra of most relevance to this study. The early data releases from ACT consisted of temperature-only data from the 2008–2010 seasons [6,23,24] using the Millimeter Bolometer Array Camera (MBAC) instrument [5,25]. MBAC was later replaced by ACTPol [7] measuring temperature and polarization using two arrays operating at 150 GHz and one dichroic array at both 98 and 150 GHz. Data from the first ACTPol seasons 2013–14 have been previously released and analyzed in Refs. [8,26]. Here we focus on ACT DR4, the 2020 release consisting of 2013–2016 ACTPol nighttime data as described in Choi *et al.* [18], henceforth C20, and Aiola *et al.* [20]. As in the C20 comparison to *Planck*, we focus here on the patches of sky comprising most of the DR4 signal to noise, "BOSS-North" (BN) and "Deep 56" (D56), illustrated in Fig. 1. These are 1837 and 565 $\mathrm{deg}^2$ patches with effective areas of 1400 and 340 $\mathrm{deg}^2$ used for power spectra after inverse-noise-variance weighting, and respective noise levels of 33–67 and $17-27$ μK-arcmin depending on array and frequency (see C20 Table 2).

### A. Ground pickup and filtering

ACT data are contaminated by substantial emission from the ground, with an amplitude of up to $\pm 20$ μK in the polarization maps [26]. Because ACT scans the sky horizontally at constant elevation this scan-synchronous ground pickup is projected into the maps as stripes of nearly constant declination. This signal is partially removed in the mapmaking step, as detailed in Aiola *et al.* [20]. Single-detector "azimuth pickup maps" that are a function of azimuth only are produced as ground templates. The CMB and other on-sky signals vary during a scan due to the rotation of the sky, and so average out over many observations. Ground pickup, however, depends only on azimuth during such a constant-elevation scan, so these azimuth maps may be used to estimate the ground signal. The azimuth pickup maps are therefore used as templates during the mapmaking procedure to subtract away the ground pickup. Despite this procedure, some residual ground pickup remains in the final frequency maps.

As in earlier analyses [8,26] this remaining signal is removed using a flat-sky Fourier-space filter removing modes $|\ell_x| < 90$ and $|\ell_y| < 50$ where $\ell = 2\pi k$ for spatial frequency $k$. The maps are cylindrical (plate carrée) projections in equatorial coordinates, so $x$ and $y$ correspond to right ascension and declination. This $k$-space filter is corrected for at the power spectrum level using a one-dimensional transfer function determined from simulations.

### B. Source-free maps

In the ACT 2013 [27] mapmaking pipeline, time-domain subtraction of point sources during the mapmaking was introduced. The motivation for this was to recover point-source fluxes better; because the beam of full width at half maximum (FWHM) 1.37 arcmin was not fully sampled by the 0.5 arcmin pixels, it was found that source fluxes measured from the maps were underestimated by about 4%. In order to recover source fluxes to 1%, the authors modeled each source in the time domain, subtracted these models from the time-ordered data (TOD), and projected them into a source-only map to be added back into the main maximum-likelihood maps. In DR4 the same TOD-level model subtraction is performed during the mapmaking for dimmer point sources. For brighter sources more care is needed; errors in the data model including sky pixelization, source variability, or gain errors can bias the maximum-likelihood mapmaker and result in leakage of signal in an "x" shape around bright point sources. These effects and mitigation strategies are studied in Ref. [28]. The approach adopted in DR4 [20] to avoid these problems is to allow an extra degree of freedom in the data model for TOD samples in the region of bright sources, jointly solving for the background sky signal and the source amplitude in each pixel. Finally, a map-based source subtraction is performed: Source locations are measured from coadded maps and a symmetric beam-shaped profile is fit to each data split and subtracted. This final step is done to remove the sources from each split map more accurately, explicitly for the purpose of avoiding artifacts from later Fourier-space filtering of the maps [20,29].

### C. Comparison to *Planck*

A direct comparison of ACT DR4 and *Planck* 2018 was carried out in C20. They computed the cross-spectra ACT × ACT (AA), ACT × *Planck* (AP), *Planck* × ACT (PA), and *Planck* × *Planck*. Note that in this notation the ordering of spectra is important, as for a TE spectrum PA uses *Planck* T and ACT E while for AP the opposite is true. The $\chi^2$ of these differences for TE and EE are given in Tables 16 and 17 of C20; direct comparisons of TT are not shown. The biggest anomalies are in TE, for which the probability to exceed (PTE) is below 0.001 for several ACT × ACT − *Planck* × ACT difference spectra (AA − PA). We reproduce selected entries of the TE table here in Table I. Here "150" and "098" refer to the ACT bands, and P is *Planck* 143 GHz. Note that those null tests involving cross-spectra at different frequencies may be complicated by the presence of foregrounds, which are expected to differ between frequency bands. This does not provide a simple explanation for the poor agreement, however, as not all of the nulls with poor PTE fall into this category, nor do all such cross-frequency nulls have poor PTE.

TABLE I. Reproduction of selected entries from Table 16 of C20, giving PTE values for differences of ACT and *Planck* cross-spectra in TE. PTEs are computed using 30 bandpowers in the range $350 \leq \ell \leq 1800$.

| D56 | PTE | BN | PTE |
|---|---|---|---|
| 150 × 150 - P × 150 | 0.000 | P × 150 - 098 × 150 | 0.000 |
| P × 098 - 098 × 098 | 0.010 | P × 098 - 098 × 098 | 0.003 |
| 150 × 098 - P × 098 | 0.017 | 150 × 150 - 098 × 150 | 0.050 |
| 150 × 098 - P × 150 | 0.017 | P × 150 - 098 × 098 | 0.070 |
| 150 × 098 - P × P | 0.027 | | |
| 150 × P - P × 150 | 0.030 | | |
| 150 × P - 098 × 150 | 0.030 | | |
| P × P - P × 098 | 0.033 | | |
| P × P - 098 × 098 | 0.033 | | |
| 150 × P - 098 × 098 | 0.043 | | |
| P × P - 098 × 150 | 0.043 | | |

The most discrepant TE spectra, with PTE $\leq 0.01$, are $\mathrm{P} \times 098 - 098 \times 098$ and $150 \times 150 - \mathrm{P} \times 150$ for D56, and $\mathrm{P} \times 098 - 098 \times 098$ and $\mathrm{P} \times 150 - 098 \times 150$ for BN. On both patches the worst PTEs are from spectra of the form $\mathrm{A}_i \times \mathrm{A}_j - \mathrm{P} \times \mathrm{A}_j$ (where $i$ and $j$ may be the same), i.e., using the same ACT polarization but exchanging ACT temperature for *Planck*. We plot the two most discrepant spectrum differences for BN in Fig. 2. Over 41 bandpowers (degrees of freedom; $N_{\mathrm{d.o.f.}}$) in the multipole range $325 < \ell \leq 2925$ we find $\hat{\chi}^2 = \chi^2/N_{\mathrm{d.o.f.}} = 1.79$ and $\hat{\chi}^2 = 1.85$ for these spectra; these correspond to $3.6\sigma$ and $3.8\sigma$ excesses. Restricting to $\ell < 1825$ as in the C20 comparison worsens the disagreement to $4.8\sigma$ and $4.1\sigma$. The discrepancy is, on the other hand, significantly improved by removing the lowest multipoles. For 25 degrees of freedom with $575 < \ell \leq 1825$ we find $\hat{\chi}^2 = 1.65$ $(2.3\sigma)$ and 1.25 $(0.9\sigma)$. Thus it would seem that the disagreement is primarily due to the differences at large scales, $\ell < 600$. This suggests that the difference may be related to the suppression of large-scale power in temperature relative to *Planck* found by C20 which led them to restrict ACT TT to $\ell > 600$ in the DR4 likelihood; this low-$\ell$ suppression, unexplained in C20, may be partially due to subpixel effects [30].

### D. *Planck* TT on ACT patches

Motivated by the above discussion, we investigate differences in the TT power spectra of ACT and an

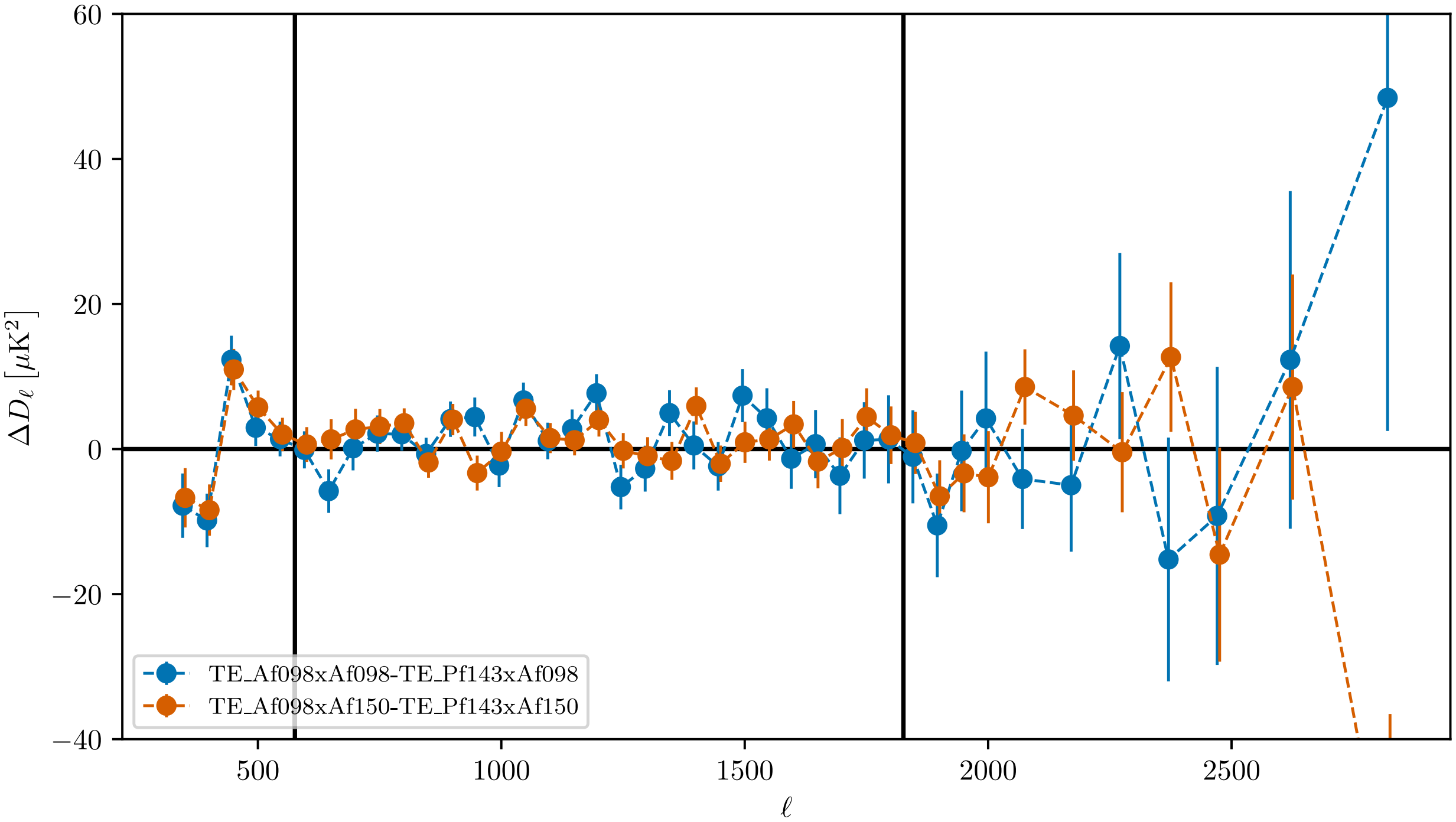

FIG. 2. The two most discrepant *Planck* × ACT − ACT × ACT TE spectra on the BN patch, as calculated in C20. The agreement is significantly improved if we restrict the $\ell$-range to the region showed by the black bars, due to improvement at low $\ell$.

ACT-like processing of *Planck* on the BN and D56 patches. We treat the *Planck* maps as follows:

(i) Rotate the maps into equatorial coordinates and project into the ACT pixelization.
(ii) Apply the Fourier-space filter introduced in Sec. II A.
(iii) Apply the BN/D56 footprint and reprojected *Planck* Galactic and point-source masks.
(iv) Estimate power spectra.

This is intended to mimic the C20 processing of the *Planck* maps, although there are small differences. In particular, we use apodized binary masks, while C20 reports using *Planck* inverse-noise-variance maps as weights. We also correct for a patch-dependent ∼1% transfer function estimated from FFP10 simulations, consistent with the expected effect of aberration on these patches.[1]

We find a systematic difference in *Planck* TT computed this way on the BN and D56 patches, illustrated in Fig. 3. This offset is due to the effect of masked point sources interacting with the filter introduced in Sec. II A. The filter spreads power from the point source across the map, far outside the point-source mask, in a crosslike pattern. If the point sources were unmasked then this would be properly accounted for by the filter transfer function, although, of course, this approach would leave significant point-source power in the computed power spectra. However, when bright sources are masked after filtering, as we shall see, the cross pattern results in the addition of a component to the power spectrum linear in $\ell$ (for $D_\ell$) and proportional to the point-source power in the map. The difference between the BN and D56 patches is therefore due to the differing point-source levels, and disappears if the maps are not filtered before point-source masking. Because this filtering primarily adds spurious power on small scales, it is not a good candidate to explain the large-scale TE discrepancies in Fig. 2; however we find the filtering effect interesting in its own right and explore it further in the following sections.

## III. ANALYTIC CALCULATION OF POINT-SOURCE CROSSES

Applying an anisotropic two-dimensional Fourier-space filter to a point source generates a crosslike pattern; in this section we derive the analytic form of these artifacts in the flat-sky approximation.

### A. Real space

We define the discrete Fourier transform and its inverse as

$$\tilde{g}_m = \mathcal{F}(g_n) = \sum_{n=0}^{N-1} g_n \exp\left(-2\pi i \frac{nm}{N}\right),$$

$$g_n = \mathcal{F}^{-1}(\tilde{g}_m) = \frac{1}{N}\sum_{m=0}^{N-1} \tilde{g}_m \exp\left(+2\pi i \frac{nm}{N}\right), \quad (1)$$

[1]Aberration [31,32] has up to a 1% level effect on the ACT patches [8,33], as BN is nearly aligned and D56 nearly anti-aligned with our direction of motion. This was corrected for in the ACT spectra of Louis *et al.* [8] and C20. The correction is not noticeable at small scales ($\ell \gtrsim 1000$) in our figures, and has no impact on our conclusions.

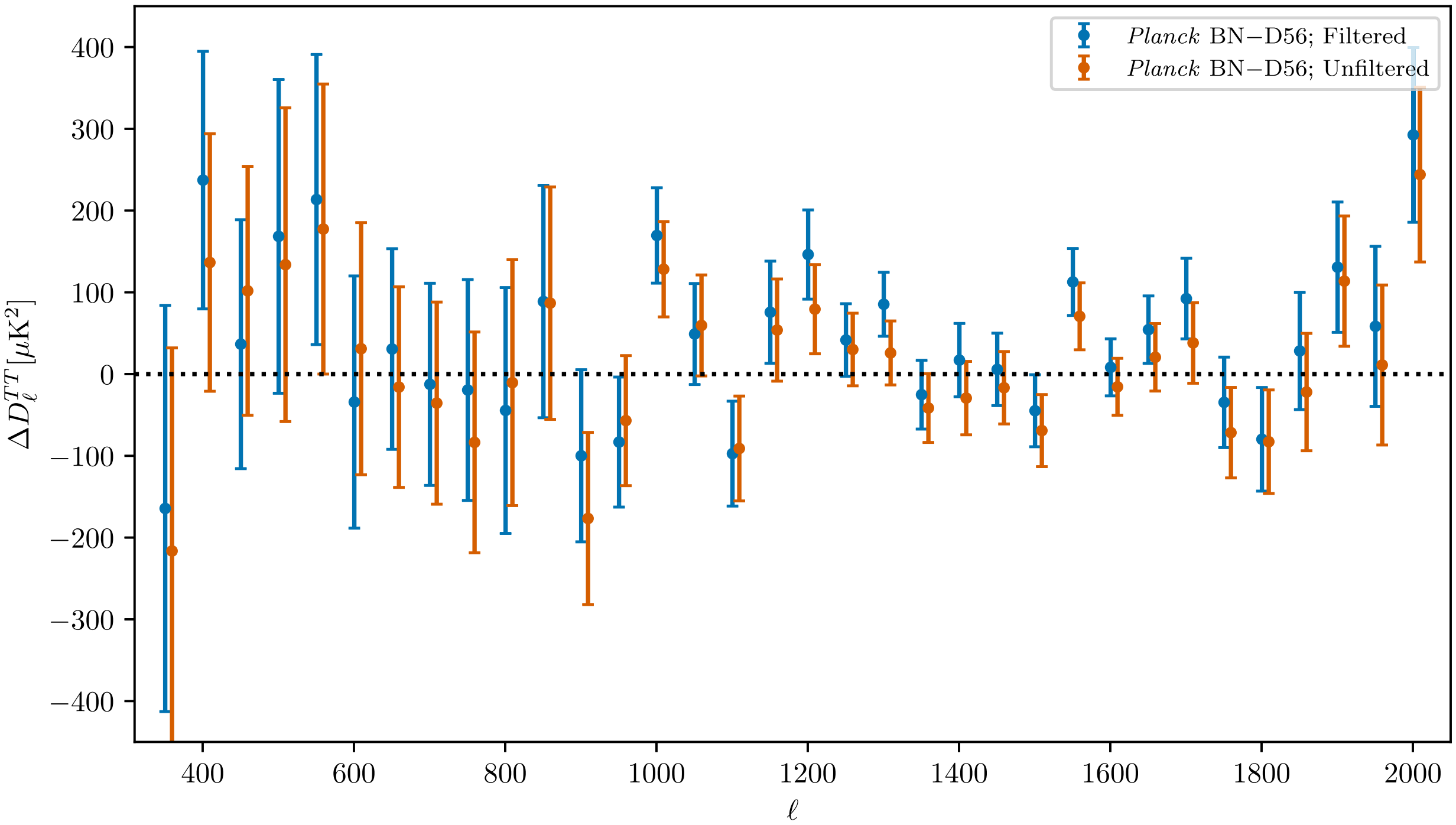

FIG. 3. Difference of *Planck* TT spectra on the BN and D56 patches for filtered (blue) and unfiltered (orange) maps. The error bars give the standard deviation of the difference computed from 300 FFP10 simulations. This difference will have contributions from foregrounds and be affected by sample variance, but the bias in the difference due to filtering is evident.

with tildes denoting Fourier-space quantities. Here, $n$ and $m$ are real and Fourier-space indices and $N$ is the total number of samples. To relate the indices to sky coordinates $x$ (rad) and spatial frequency $k$ (rad$^{-1}$) we define the real pixel width $d_x$ and the step size in $k$, $d_k = 1/W$, where $W$ is the full width of the map. Then $x_n = d_x n + x_0$ and $k_m = d_k m$ for $m < N/2$ and $k_m = d_k(m - N)$ for $m \geq N/2$. Now for a map $M$ and a Fourier-space filter $\tilde{f}$ the filtered map $\overline{M}$ is given as

$$\overline{M} = \mathcal{F}^{-1}(\tilde{f} \times \mathcal{F}(M)) = \mathcal{F}^{-1}(\tilde{f}) * M. \tag{2}$$

Here, $\mathcal{F}^{-1}$ denotes an inverse Fourier transform and $*$ a convolution.

For the application considered here, $\tilde{f}$ is essentially two boxcar functions $b$ at right angles to each other,

$$\tilde{f} = b_x(k_x) b_y(k_y). \tag{3}$$

We define $b(k) = 1 - \Pi(\frac{k}{2k_f})$, where $\Pi(x)$ is the rectangular function equal to 1 for $|x| < \frac{1}{2}$ and 0 elsewhere. The half width of the filter is $k_f$; in ACT DR4 $k_{fx} = 90/2\pi$ and $k_{fy} = 50/2\pi$ where we use $k = \ell/2\pi$ in the flat-sky approximation. We define $m_f$ as the index of the largest $k$ value below $k_f$, that is $m_f = \lfloor k_f/d_k \rfloor$. Taking the inverse Fourier transform of $\tilde{f}_k$ in one dimension we find

$$f_n = \delta_{n0} - \frac{1}{N} \frac{\sin\left(\frac{\pi n}{N}(2m_f + 1)\right)}{\sin(\frac{\pi n}{N})}, \tag{4}$$

where $\delta_{n0}$ is the Kronecker $\delta$.

In one dimension with a $\delta$-function point source, we have an aliased sinc function that will extend far outside the point-source mask, spreading the point-source power across the map; in two dimensions, these are the "crosses." This is straightforwardly extended to two dimensions. Convolving with a point source described with a beam profile $B$ (normalized to unit area) and a flux $\varphi$, we get the cross profile:

$$\mathrm{Cross}(\mathbf{x}) = \varphi B(\mathbf{x}) * \left(\left[\delta_{n_x 0} - \frac{1}{N_x} \frac{\sin\left(\frac{\pi n_x}{N}(2m_{fx} + 1)\right)}{\sin(\frac{\pi n_x}{N})}\right] \times \left[\delta_{n_y 0} - \frac{1}{N_y} \frac{\sin\left(\frac{\pi n_y}{N}(2m_{fy} + 1)\right)}{\sin(\frac{\pi n_y}{N})}\right]\right). \tag{5}$$

Such a pattern is shown in Fig. 4. We also give the slightly simpler expression using a continuous instead of discrete Fourier transform, accurate far from the map edges (we match the normalization to the discrete case above):

$$\mathrm{Cross}(\mathbf{x}) \approx \varphi B(\mathbf{x}) * \left(\left[\delta(x) - \frac{d_x}{\pi} \frac{\sin(x\hat{\ell}_{fx})}{x}\right] \times \left[\delta(y) - \frac{d_y}{\pi} \frac{\sin(y\hat{\ell}_{fy})}{y}\right]\right), \tag{6}$$

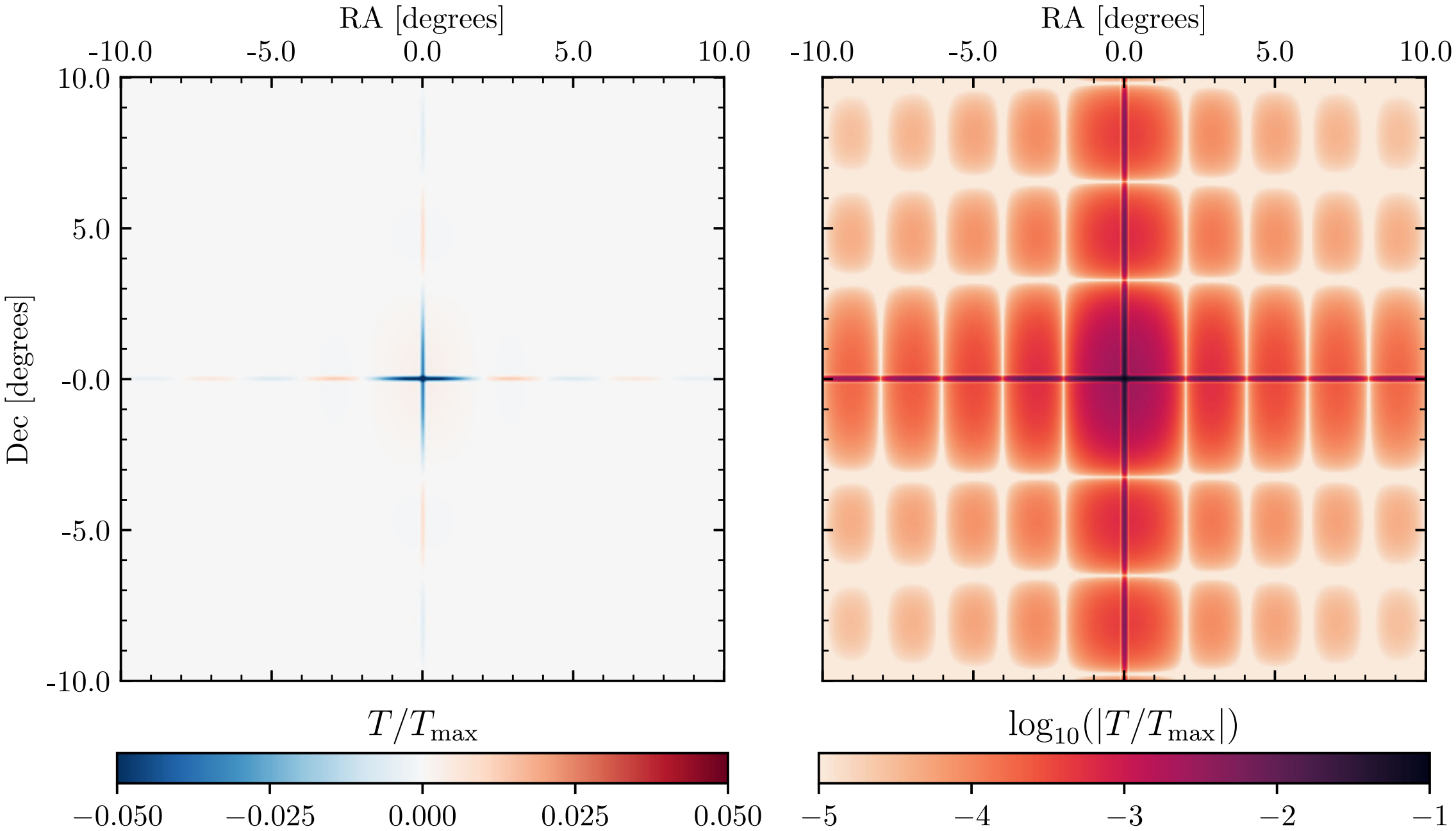

FIG. 4. "Cross" pattern appearing when the ACT $k$-space filter is applied to a point source, here a 7 arcmin FWHM Gaussian, which is subtracted out after filtering. The color scale shows $T/T_{\rm max}$ where $T_{\rm max}$ is the peak pixel value of the unfiltered source. The left panel saturates at the center, with peak $T/T_{\rm max} \approx 0.1$; the right panel shows the same image on a logarithmic scale for more detail. A 1 Jy source in *Planck* 143 GHz has a maximum of 510 μK, so the absolute value of the cross peaks around 50 μK.

where the center of the map is taken to be at (0,0) and $\hat{\ell}_f = (m_f + \frac{1}{2})d_\ell$ is the effective largest $\ell$ cut.

### B. Power spectrum

Now we want to calculate the effect of the cross pattern on the angular power spectrum. If we assume that the point source is masked after the whole map is filtered then we are approximately measuring the power spectrum of

$$\overline{S+M} - S = (\overline{S} - S) + \overline{M} = X + \overline{M}. \tag{7}$$

That is, we approximate the point-source masking by subtraction. Here, $S$ is the point source and $M$ the remainder of the map (CMB, noise), both convolved by the beam. As above, an overline denotes a filtered quantity so $\overline{S} = \mathcal{F}^{-1}(\tilde{f} \times \mathcal{F}(S))$ and we define $X \equiv \overline{S} - S$. We ignore the small portion of the CMB that gets masked along with the point source. We can calculate the power spectrum of the sum $X + \overline{M}$ as

$$C^{(X+\overline{M})\times(X+\overline{M})} = C^{XX} + C^{X\overline{M}} + C^{\overline{M}X} + C^{\overline{M}\overline{M}}. \tag{8}$$

The cross terms will add some variance but are zero on average since the cross pattern is uncorrelated with the CMB; we return to this extra variance in Sec. IV, now focusing on $C^{XX}$ and $C^{\overline{M}\overline{M}}$. We calculate these in the flat-sky approximation, starting with the empirical two-dimensional power spectrum of the cross pattern $X$:

$$\begin{aligned}\hat{P}_{2D}(X) &= \mathcal{F}(X)\mathcal{F}(X)^* \\ &= [\tilde{f}\tilde{S} - \tilde{S}][\tilde{f}\tilde{S} - \tilde{S}]^* \\ &= \tilde{S}\tilde{S}^*(\tilde{f}\tilde{f}^* - \tilde{f} - \tilde{f}^* + 1).\end{aligned} \tag{9}$$

Noting that $\tilde{f}$ only takes the values 0 and 1, $\tilde{f} = \tilde{f}^* = \tilde{f}\tilde{f}^*$, the term in parentheses in the last line simplifies to $1 - \tilde{f}$. To calculate the one-dimensional (angular) power spectrum we average over azimuthal angles, and assume $\tilde{S}\tilde{S}^*$ is azimuthally symmetric. Rewriting $\tilde{S}\tilde{S}^* = P(S)$ and denoting the azimuthal average of $\tilde{f}$ as $\tilde{f}_{\langle\theta\rangle}$ we have

$$P(X) = (1 - \tilde{f}_{\langle\theta\rangle})P(S). \tag{10}$$

This expression holds similarly for the expected power spectrum of the crosses from a population of unclustered point sources, whose underlying power spectrum is $P(S)$.

Repeating this calculation for the map, the empirical 2D power spectrum is

$$\begin{aligned}\hat{P}_{2D}(\overline{M}) &= [\tilde{f}\tilde{M}][\tilde{f}\tilde{M}]^* \\ &= \tilde{M}\tilde{M}^*\tilde{f}.\end{aligned} \tag{11}$$

Taking the mean over realizations of the map, which is assumed to be statistically isotropic with power spectrum $P(M)$, we obtain the expected angular power spectrum

$$P(\overline{M}) = \tilde{f}_{\langle\theta\rangle}P(M). \tag{12}$$

To estimate the desired power spectrum $P(M)$ we will divide out by the filter transfer function, which we can see is given by $\tilde{f}_{\langle\theta\rangle}$. Applying this operation we get the transfer-function-"corrected" power spectrum of the crosses

$$C^{XX} = \frac{1 - \tilde{f}_{\langle\theta\rangle}}{\tilde{f}_{\langle\theta\rangle}} P(S). \tag{13}$$

We now calculate $\tilde{f}_{\langle\theta\rangle}$:

$$\tilde{f}_{\langle\theta\rangle} = \frac{1}{2\pi}\int_0^{2\pi} \mathrm{d}\theta \tilde{f}. \tag{14}$$

If $\ell \leq \sqrt{\ell_{fx}^2 + \ell_{fy}^2}$ then $\tilde{f}_{\langle\theta\rangle}(\ell) = \tilde{f}(\ell, \theta) = 0$. Otherwise we find

$$\tilde{f}_{\langle\theta\rangle}(\ell) = 1 - \frac{2}{\pi}\left[\arcsin\left(\frac{\ell_{fx}}{\ell}\right) + \arcsin\left(\frac{\ell_{fy}}{\ell}\right)\right]. \tag{15}$$

Note that in the discrete case $\ell_f$ should be replaced by the upper edge of the highest "$\ell$-pixel" filtered, $\hat{\ell}_f$, as above.

We are primarily interested in large $\ell$ so we expand $C^{XX}$ to second order in $1/\ell$ (i.e., neglecting terms that decay in $D_\ell$) and defining $\hat{\ell}_f = \hat{\ell}_{fx} + \hat{\ell}_{fy}$ we find

$$C_\ell^{XX} = P(S)\left[\frac{2}{\pi}\frac{\hat{\ell}_f}{\ell} + \left(\frac{2}{\pi}\frac{\hat{\ell}_f}{\ell}\right)^2\right], \tag{16}$$

$$D_\ell^{XX} = P(S)\frac{\hat{\ell}_f}{\pi^2}\left[\ell + \left(\frac{2}{\pi}\hat{\ell}_f + 1\right)\right]. \tag{17}$$

For ACT DR4 values of $\ell_f$ this expansion is correct to 10% at $\ell > 300$ and 1% at $\ell > 950$.

The power spectrum $P(S)$ of an isotropic population of point sources is, after beam deconvolution, constant in $\ell$ (see, e.g., Tegmark and Efstathiou [34]). We see that this constant source power leaks into the final power spectrum with an extra $1/\ell$ suppression factor, which turns into a linear trend in $\ell$ in $D_\ell$. For bright sources this contribution can be very substantial.

## IV. IMPACT

In this section we discuss the effects of filtering of point sources with the example case of ACT DR4, particularly the BN patch. We focus on BN as it has somewhat more point-source power than D56. If we estimate the source power on a given patch by computing the mean square of the source-only map, then we find the ratio BN/D56 to be 2.2. As we have seen above and confirm later in this section, the offset due to the crosses scales with the source power, and so we expect a larger effect in BN. Indeed this can be directly observed in a comparison of the *Planck* TT estimates from each patch. As illustrated in Fig. 3, the BN spectrum is greater than D56 by around 20–40 μK$^2$ in $D_\ell$, although there is significant scatter in the difference due to sample variance. This offset is only present when the maps are filtered before point-source masking, and arises due to the difference in point-source power. Such a difference between power spectra as calculated on different parts of the sky is one way the impact of the crosses can manifest; we further explore their effects on the BN temperature and polarization power spectra below.

### A. ACT-filtered *Planck* TT

We begin with the case of *Planck* TT; since there are substantially fewer sources in polarization and ACT uses source-subtracted maps this is the only case in which we see a strong effect from the crosses, as will be verified and discussed further in Sec. IV B.

In Fig. 5 we see the results of applying the ACT DR4 $k$-space filter with $\ell_{fx} = 90$ and $\ell_{fy} = 50$ to the *Planck* data on the BN patch. Here we use the *Planck* 2018 half-mission maps and masks from the Planck Legacy Archive.[2] The difference of the spectra of filtered and unfiltered *Planck* maps (with the source mask subsequently applied in both cases), $\overline{\mathrm{P}}_1 \times \overline{\mathrm{P}}_2 - \mathrm{P}_1 \times \mathrm{P}_2$ where $\mathrm{P}_1$ and $\mathrm{P}_2$ indicate *Planck* 143 GHz half-missions 1 and 2, respectively, is shown in blue. Referring back to Eq. (8), this measurement of the effect of filtering on the *Planck* spectrum should be equivalent to the power spectrum of the crosses $C^{XX}$ [Eq. (13) after transfer-function deconvolution] in the limit that the cross terms $C^{\bar{M}X}$ average to zero, as expected.

As is visible in the figure there is substantial variance on this difference; this comes primarily from the modes of the map $M$ that are removed from the filtered spectrum but are included on the unfiltered side. In the approximation that each mode $a_{\ell m}$ is either completely filtered or left unchanged,[3] the filter removes a fraction $1 - t_f(\ell)$ of modes, where $t_f = \tilde{f}_{\langle\theta\rangle}$ is the same as the filter transfer function derived in Sec. III. Then we have the difference of the (transfer-function-corrected) empirical power spectrum of the filtered and unfiltered maps

$$\hat{C}_\ell^{\overline{MM}} - \hat{C}_\ell^{MM} = \frac{1}{\nu}\left(\sum_{m_1}\left(\frac{1}{t_f} - 1\right) a_{\ell m_1} a^*_{\ell m_1} - \sum_{m_2} a_{\ell m_2} a^*_{\ell m_2}\right), \tag{18}$$

where $\nu = (2\ell + 1) f_{\mathrm{sky}}$ is the usual number of modes on a sky fraction $f_{\mathrm{sky}}$, the first sum is over $\nu t_f$ unfiltered modes $m_1$, and the second is over $\nu(1 - t_f)$ filtered modes $m_2$.

[2] http://pla.esac.esa.int/pla/

[3] The modes here should be understood to be discrete flat-sky Fourier modes, for which the filtered and unfiltered modes are uncorrelated.

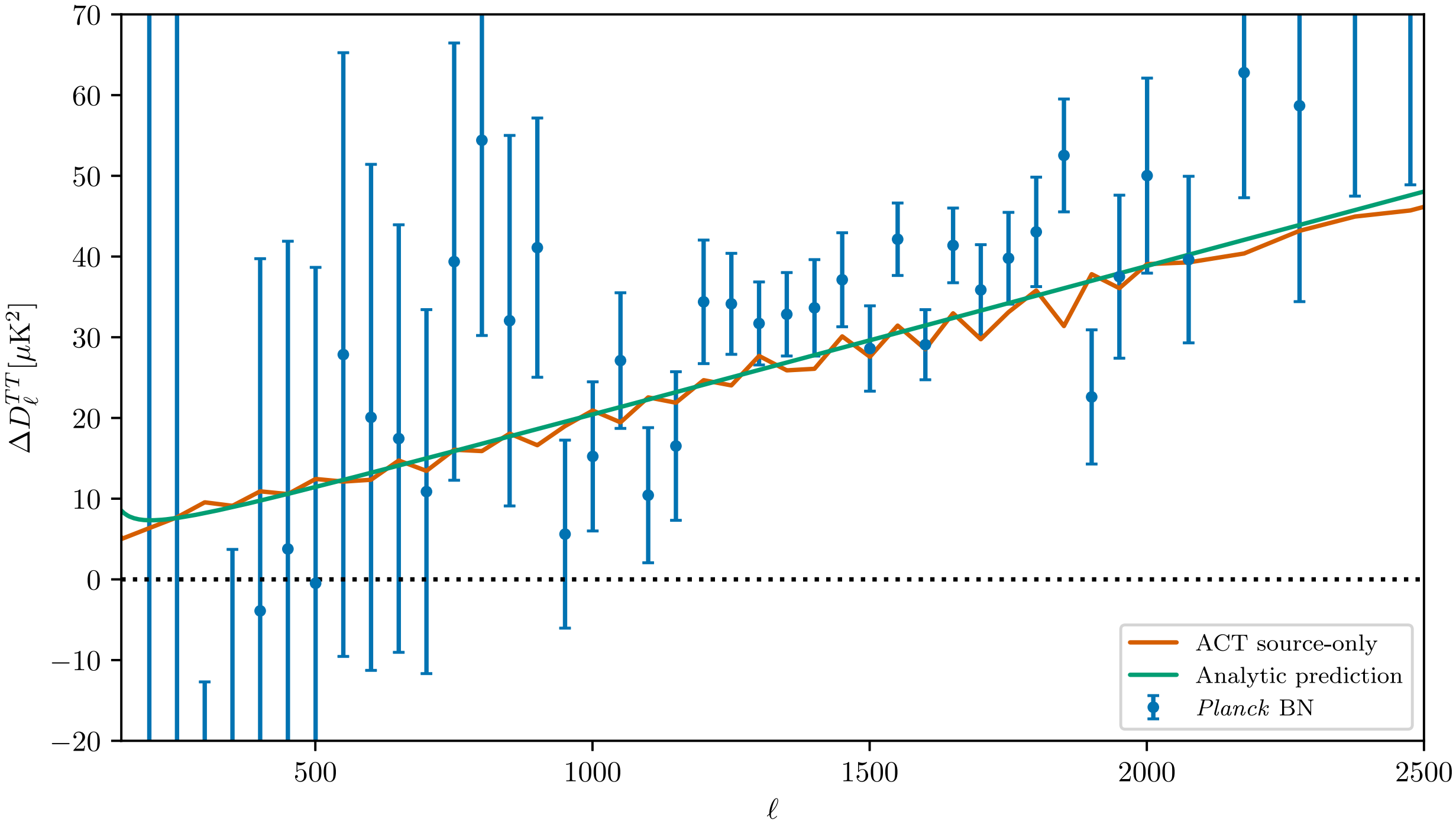

FIG. 5. The effect of filtered point sources on the power spectrum $D_\ell$, on the BN patch. In blue we show the difference of *Planck* power spectra for maps that are filtered or not filtered before applying the point-source mask. The orange line shows the power spectrum of a filtered and masked source-only map from ACT, with good agreement to the *Planck* measurement. The green line shows the analytic prediction for the power spectrum of the crosses from Eq. (22).

The expectation value of $\hat{C}_\ell^{\overline{MM}} - \hat{C}_\ell^{MM}$ is zero, but there is a nonzero variance given in the noiseless case by

$$\sigma^2 = \langle(\hat{C}_\ell^{\overline{MM}} - \hat{C}_\ell^{MM})^2\rangle = \frac{2(C_\ell^{MM})^2}{\nu}\left(\frac{1}{t_f} - 1\right)$$
$$= \frac{\sigma_{\mathrm{CV}}^2}{f_{\mathrm{sky}}}\left(\frac{1}{t_f} - 1\right). \tag{19}$$

To include noise we simply replace the cosmic variance $\sigma_{\mathrm{CV}}^2$ with the diagonals $\sigma_P^2$ of the binned covariance matrix for the *Planck* $143 \times 143$ spectra, and rescale to the BN patch by a ratio of $f_{\mathrm{sky}}$ factors.[4] That is,

$$\sigma = \left[\sigma_P^2 \frac{f_{\mathrm{sky}}^{\mathrm{P}}}{f_{\mathrm{sky}}^{\mathrm{BN}}}\left(\frac{1}{t_f} - 1\right)\right]^{1/2}. \tag{20}$$

This assumes isotropic noise among other simplifications, but is sufficient for our purposes. We compare this analytic error estimate to that computed from simulations, in particular the *Planck* FFP10 simulations, on which we calculate the same filtered-unfiltered difference. We find good agreement between the simulations and Eq. (20), although our analytic expression somewhat underestimates the error relative to the simulations at $\ell > 1500$. For the error bars in Fig. 5 we use those from the simulations.

Still referring to Fig. 5, we compare the *Planck* difference to the spectrum of filtered sources from ACT in orange. Here we use ACT DR4 source-only maps to measure the power spectrum of the crosses directly, without needing to subtract the CMB contribution. The source maps consist of ACT beam profiles with amplitudes fit in map space for all sources detected at SNR > 5 in optimally coadded single-frequency maps [20]. The ACT source maps contain some sources that are not included in the *Planck* mask due to the higher resolution and greater sensitivity of ACT. We do not want to include such sources that are not masked by *Planck* here, as they should be correctly deconvolved by the filter transfer function, and so should simply add the same point-source power to both the filtered and unfiltered maps. We remove these unmasked sources by applying the (additive) inverse of the *Planck* source mask to the source map. We then apply the $k$-space filter to generate the crosses as usual, and then apply the *Planck* source mask. Denoting the *Planck* source mask as $W_S$ and the ACT source map as $S_{\mathrm{ACT}}$, we are calculating $P(W_S \times \mathrm{Filt}((1 - W_S)S_{\mathrm{ACT}}))$, where Filt denotes the $k$-space filtering operation and $P$ the power spectrum. As expected this approach gives nearly identical results to the difference of the filtered and unfiltered source maps.

[4] In fact, we account for apodization by using the expression $(w_2^2/w_4)f_{\mathrm{sky}}$ [35]. Here $f_{\mathrm{sky}}$ is the fraction of sky area with nonzero weight and $w_i$ is the $i$th moment of the mask over that area.

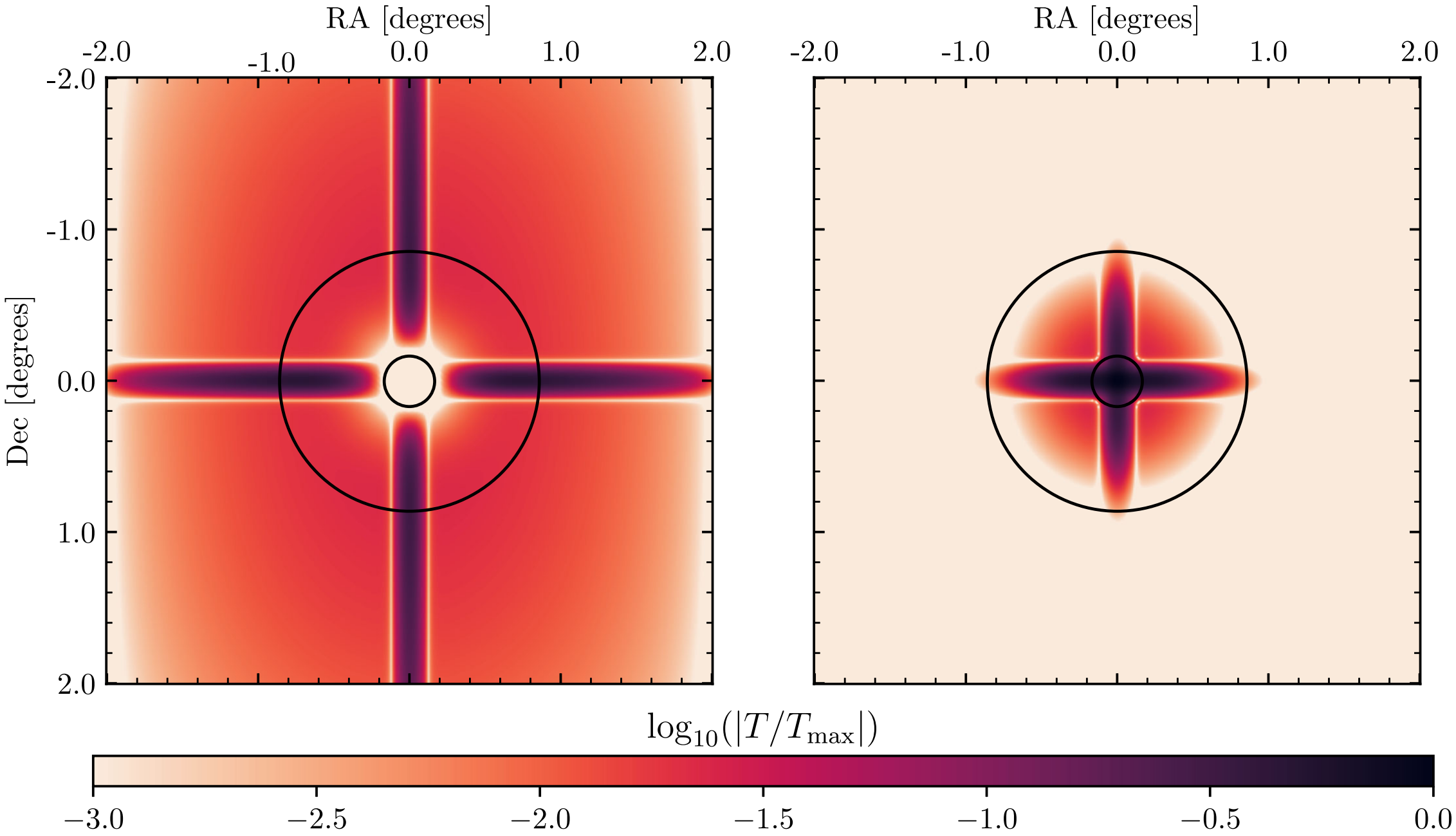

FIG. 6. Left: the cross of Fig. 4 enlarged and with a *Planck* 143 GHz source mask applied. Right: the portion of the cross that has been removed by the (apodized) mask. The color scale is relative to the maximum of the unmasked cross. The inner and outer black circles indicate where the apodized mask is equal to 0 and 0.99, respectively.

We recover the expected linear trend in $D_\ell$ [Eq. (17)] with a very good match to the *Planck* difference.

Finally, in Fig. 5 we also plot the theoretical prediction from Eq. (13) in green. The amplitude is given by the power spectrum of the source map $P(S_{\rm ACT})$, where, now, we implicitly include multiplication by the inverse source mask from *Planck* in $S_{\rm ACT}$. The power spectrum of an individual point source of flux $\varphi$ is constant, $C_\ell = \varphi^2/(4\pi)$; for a population of randomly distributed sources the (theoretical) power spectrum $\langle C_\ell \rangle$ remains a constant related to the sum of the squared fluxes. The expression is given by Tegmark and Efstathiou [34]:

$$C_\ell = \int_0^{\varphi_c} \frac{\partial \bar{n}}{\partial \varphi} \varphi^2 {\rm d}\varphi, \tag{21}$$

where $\bar{n}(\varphi)$ is the number density per steradian of sources with flux less than $\varphi$. We do not apply Eq. (21) directly but rather measure this constant $C_\ell = A_{\rm src}$ empirically from the ACT source map. However, using this number directly in Eq. (13) overpredicts the power spectrum of the crosses by a factor of approximately 1.8. This is because we have calculated the power from the full cross (after source subtraction), but a significant additional amount from the central region is removed by the point-source mask as illustrated in Fig. 6. We account for this effect by computing the ratio of power spectra of crosses with and without the mask, $C_{\rm mask} = \langle \frac{P(W(\tilde{S}-S))}{P(\tilde{S}-S)} \rangle_\ell$, where the angle brackets denote flat averaging in the multipole range 1000–2500, leading to

$$C_\ell^{XX} = C_{\rm mask} A_{\rm src} \frac{1 - t_f}{t_f}. \tag{22}$$

For the present case we use the coadded source map from ACT 150 GHz, season 15 and array PA1, and find $A_{\rm src} = 0.00205\ \mu{\rm K}^2$ and $C_{\rm mask} = 0.57$. With this correction, we again see good agreement with the direct *Planck* and ACT measurements of the power spectrum of the crosses. Figure 5 therefore offers empirical verification that the crosses coming from either ACT or *Planck* point sources have a linear contribution to $D_\ell$, in agreement with Eq. (17). In this case we find $D_\ell^{\rm bias} \approx (0.018\ell + 1.7)\ \mu{\rm K}^2$; the amplitude is up to about $30\ \mu{\rm K}^2$ by $\ell = 1500$, amounting to a significant bias in the BN patch. Applying the same analysis for D56 at 150 GHz, we find $D_\ell^{\rm bias} \approx (0.007\ell + 0.6)\ \mu{\rm K}^2$, an $11\ \mu{\rm K}^2$ bias at $\ell = 1500$.

### B. TE and *Planck* × ACT

If the masked sources are very few or weak then $A_{\rm src}$ of Eq. (22) will be very small. Therefore if point sources in polarization are negligible, as suggested by, e.g., Efstathiou and Gratton ( [36], Sec. 3. 2) and Naess *et al.* [26], then the bias explored in the previous section should not be present in the TE or EE spectra. This is illustrated in Fig. 7, where similar to the blue points of Fig. 5 we compare power spectra of filtered and unfiltered *Planck* maps (both subsequently masked), now for TE. The additional power evident in TT is absent for both patches in TE, confirming our expectation. This also shows that bias due to filtering of point sources cannot explain the high $\chi^2$ values reported in C20 and discussed in Sec. II.

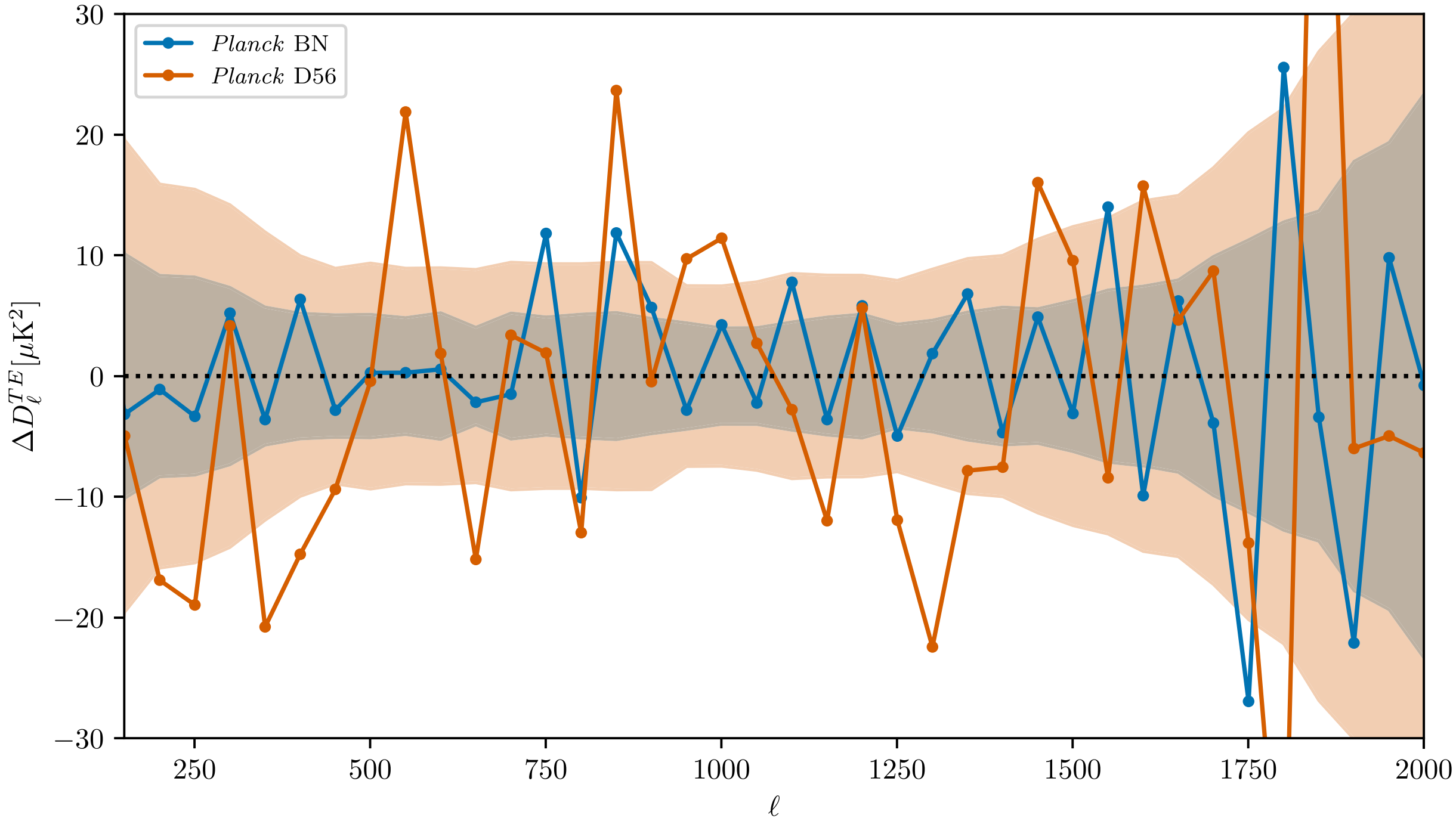

FIG. 7. Differences of TE power spectra of filtered and unfiltered *Planck* maps (subsequently masked with the *Planck* point-source mask in both cases) for the BN (blue) and D56 (orange) patches. The colored bands show the $1\sigma$ error region as determined from simulations.

We also check if any significant bias is incurred by crosses induced in the ACT source-subtracted maps by source residuals. Such residuals can appear in both temperature and polarization; a particularly strong example from BN is shown in Fig. 8. The application of the filter does induce a cross pattern, though it is weaker and somewhat different in detail from the usual case due to the lower amplitude and different shape of the source residual. We could test for this effect with a difference of filtered and unfiltered spectra as above, but as we have seen this method is limited by variance coming from the filtered modes. Therefore to estimate the bias incurred by these residuals-induced crosses more precisely, we calculate the power spectrum $P(W_S \times \mathrm{Filt}((1 - W_S)A_{SF}))$, where, again, $W_S$ is the point-source mask and now $A_{SF}$ is an ACT source-subtracted map. That is, we apply the inverse source mask to remove the majority of the CMB and only calculate the power spectrum inside the source masks, where we expect possible source-subtraction residuals. As the source residuals tend to be faint, we minimize the impact of CMB on the calculation by subtracting a mean value around each point and masking more aggressively. We calculate the background from an annulus around each point, of inner radius 4.5 arcmin and outer radius 7 arcmin, and mask the annulus and exterior. We find negligible power from these residuals. We use the same method to check if there may be a larger effect from cross-correlating these source residuals with the filtered sources, as in the *Planck* case, and again find the bias to be negligible. We note that the absence of bias in *Planck*×ACT spectra implies that the ACT TT calibration off of the $A \times A - P \times A$ difference is unaffected by this effect.

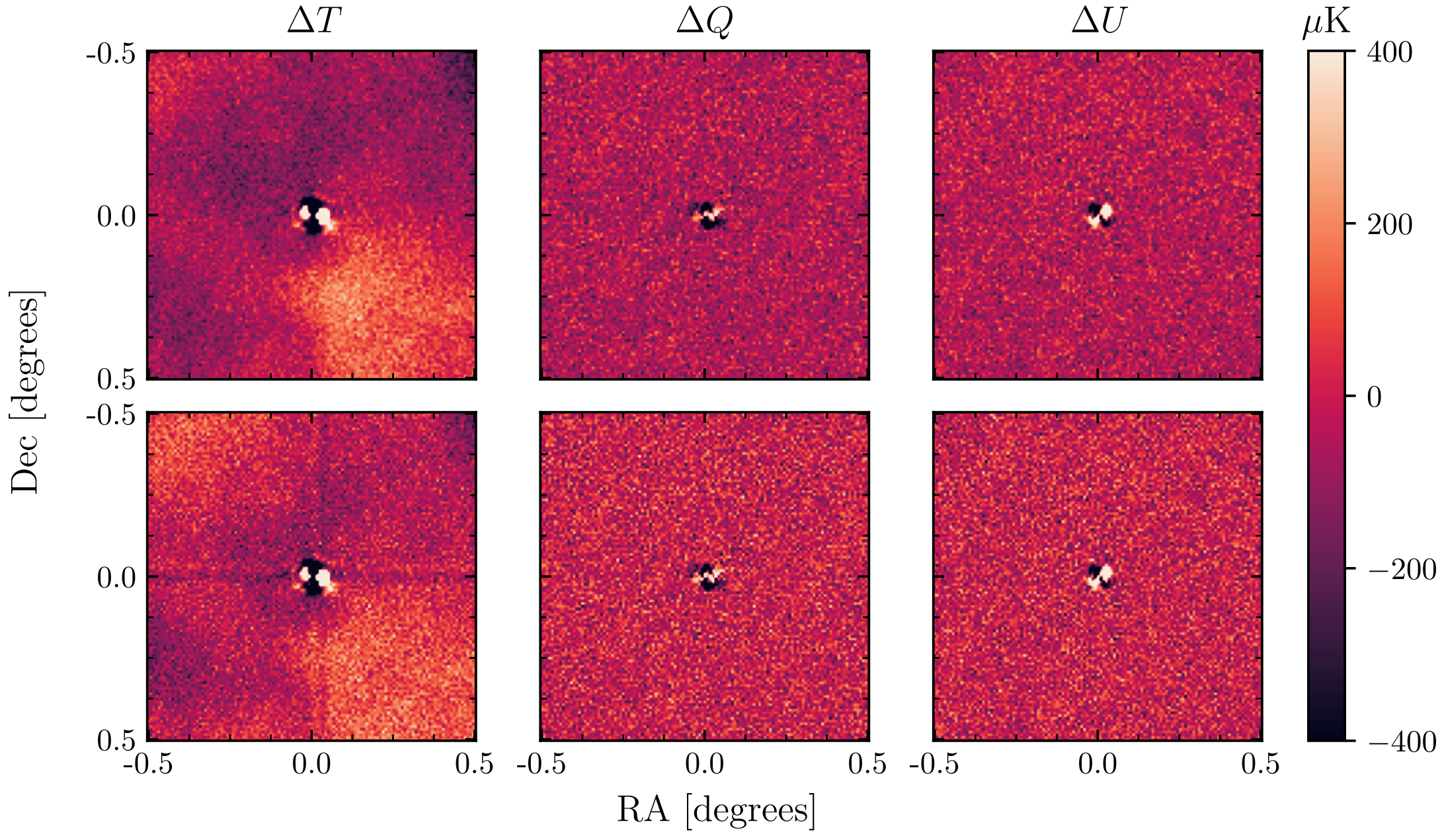

FIG. 8. Example of residuals after source subtraction in the BN patch. The columns show $T$, $Q$, and $U$ in μK. The top row is as in the ACT source-subtracted maps, and the lower after application of the $k$-space filter; a faint cross is visible in temperature.

Having established that the bias due to crosses is small in TE and when using source-subtracted maps, we finally check the magnitude of the additional variance coming from the Cross ×CMB terms, as mentioned in Sec. III. To study this additional variance in more detail we use simple simulations to compute the typical amount of extra scatter. We make a representative cross pattern as before by filtering the ACT source map after multiplying by the inverse point-source mask from *Planck* to select those sources masked by *Planck*. We generate 100 simple CMB-only simulations on the BN patch, using the *Planck* 2018 best fit [37] as our fiducial cosmology. We add the same cross pattern $X$ to each CMB map and compute $X \times \mathrm{CMB}$ for TT and TE. We compute the standard deviation $\sigma$ of these simulated spectra,[5] and compare to the *Planck* × ACT − ACT × ACT difference as computed for C20. We find approximately 0.2 $\mu\mathrm{K}^2$ of additional scatter in TE, an order of magnitude lower than the error bars from C20. While this is a simple noiseless treatment, the inclusion of noise would not change the conclusion that this extra variance is far from explaining the high $\chi^2$ of the PA − AA difference.

## V. MITIGATION

We have seen that $k$-space filtering of point sources leads to leakage of their power to the rest of the map, which is then not masked by usual point-source masks. In the example case of *Planck* on the BN patch we find that this leads to a substantial bias in TT, whereas for TE and *Planck* ×ACT it adds slightly to the variance. This extra variance is small in our case, but could be important at higher levels of precision or for other patches with stronger point sources. In our example case these issues can, of course, be avoided by simply not filtering the *Planck* data. It may however be desirable to process *Planck* data in a similar fashion to ACT, and, as discussed above, filtering is generally necessary for ground-based experiments. We therefore introduce here methods for mitigation of this error; related methods are also discussed, from a mapmaking perspective, in Refs. [20,28].

### A. Masking and in-painting

If subtraction of sources is not possible (e.g., there is no catalog with reliable fluxes), the most straightforward method of removing the effect of sources is to mask them before applying the Fourier-space filter. This obviously does not work for applications when sources need to be retained in the ground-cleaned map, but for a straightforward primary CMB analysis it would be a simple method of removing the artifacts due to the point sources. Application of a complicated mask before performing Fourier-space operations could introduce ringing in the map; however given that this would be the same (apodized) mask as applied before normal power spectrum computation, there should be no additional ringing effect. On the other hand, this method could introduce some dependence of the filter transfer function on the point-source mask. Worse, mode coupling introduced by the source mask could significantly break the assumption that the filter can be corrected with a one-dimensional transfer function. Using simulations of the BN patch we found this method to be effective and unbiased in our present case, and significant effects on the filter transfer function were not detected. When applying the method to other patches or datasets, however, we recommend further testing to confirm that these results continue to hold to the desired precision.

An alternative to masking the sources is in-painting, one implementation of which is described in Bucher and Louis [38] and was adopted for the ACT DR4 tSZ analysis [29]. In this method, detected sources are masked as previously, but the resulting holes are filled with a correlated random field with the same statistics as the sky signal (before filtering). This approach therefore mitigates any concerns about ringing as well as the mask dependence of the transfer function. It does require accurate reproduction of the statistical properties of the local sky and noise in temperature and polarization, and is therefore significantly more complicated than simple masking. This increased complexity should not pose a significant problem however, as effective in-painting algorithms such as “constrained Gaussian realizations” [38] have already been implemented.

[5]Note that for TT this $\sigma$ is as for *Planck* ×ACT, and would be doubled if both maps had crosses as for *Planck* × *Planck*.

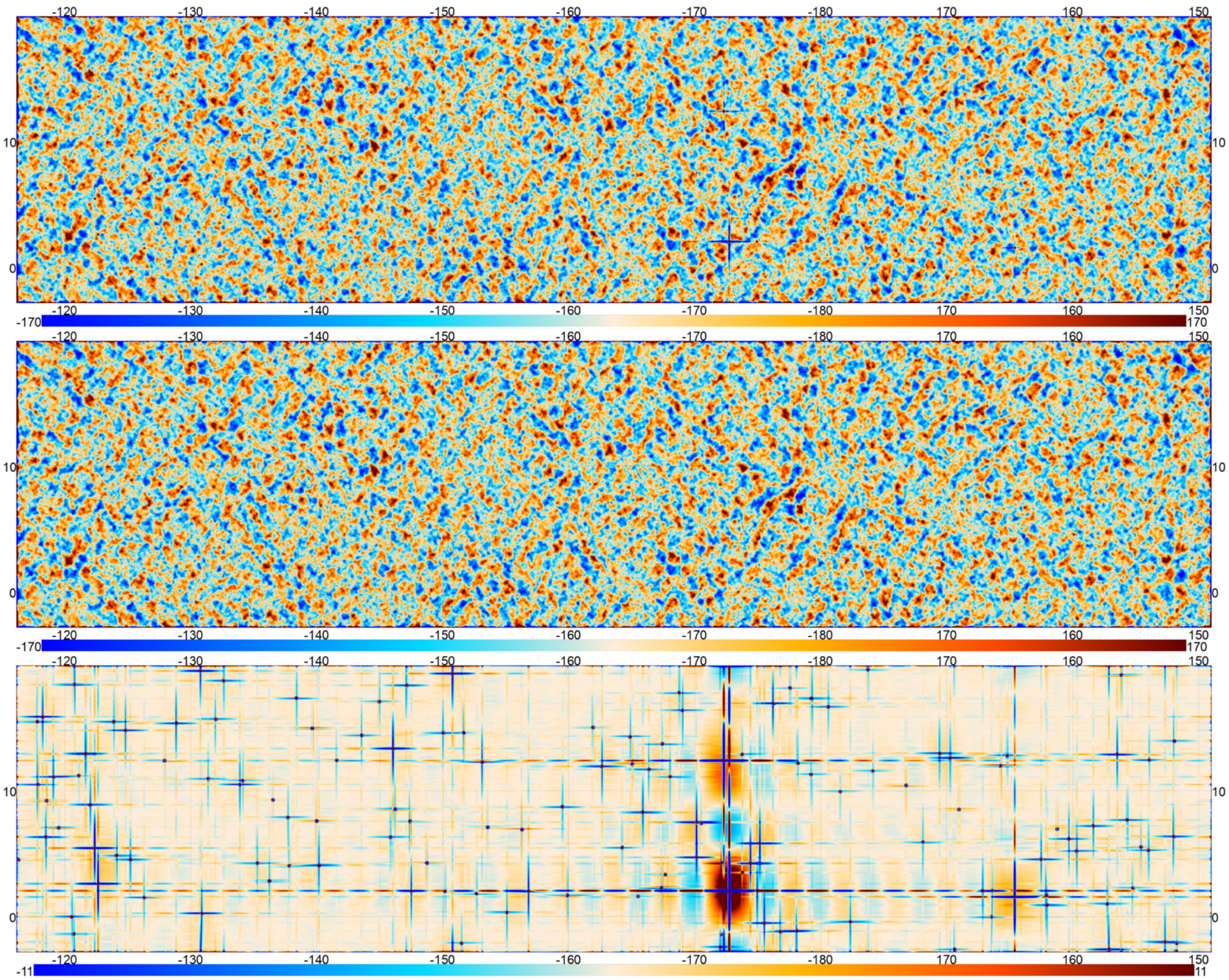

FIG. 9. Filtered *Planck* maps on the BN patch. The top row is with no in-painting. In the second row, sources in the *Planck* point-source catalog have been in-painted before filtering. The third panel is their difference. The map coordinates are right ascension and declination in degrees, and the color scale is in units of μK. The extended red and blue patches and horizontal/vertical stripes in the bottom panel are due to particularly strong sources at (RA, Dec) of (−173, +3) and (−172, +13); compare to Fig. 4.

We illustrate in-painting using *Planck* 143 GHz maps on the BN patch in Fig. 9. The top panel shows a filtered map with evident cross patterns around the brightest sources. The second panel has been in-painted before filtering, eliminating these patterns; in particular we in-fill 135 holes of 9 arcmin radius at locations defined in the *Planck* source catalog. The third panel, showing the residual map, illustrates the contaminating crosses that have been eliminated by the in-painting procedure, including many that were not obvious in the first panel. In Fig. 10 we show the effect on power spectra by recreating Fig. 5, now in-painting the *Planck* maps before applying the Fourier-space filtering. The filtered-unfiltered power spectrum difference in Fig. 10 is consistent with null, with a reduced $\chi^2$ of 0.9 for 43 bandpowers between $\ell$ of 150 and 2500.

### B. Source subtraction

Finally, another alternative to masking is to fit and subtract out sources from the map, as done for ACT DR4 [20]. In the implementation presented there, the usual source catalog is constructed using a matched filter on coadded maps. Source amplitudes are fit on individual maps, with profiles identical to the beam, and subtracted from each map. Imperfect subtraction due to asymmetric beams or closely blended sources can leave residuals in the map, as was shown in Fig. 8; however, as shown on BN in Sec. IV B, the effect of these residuals on the power spectrum is generally small. As we have seen, even partial or incomplete subtraction can significantly decrease the bias since the cross effect scales with the point-source power.

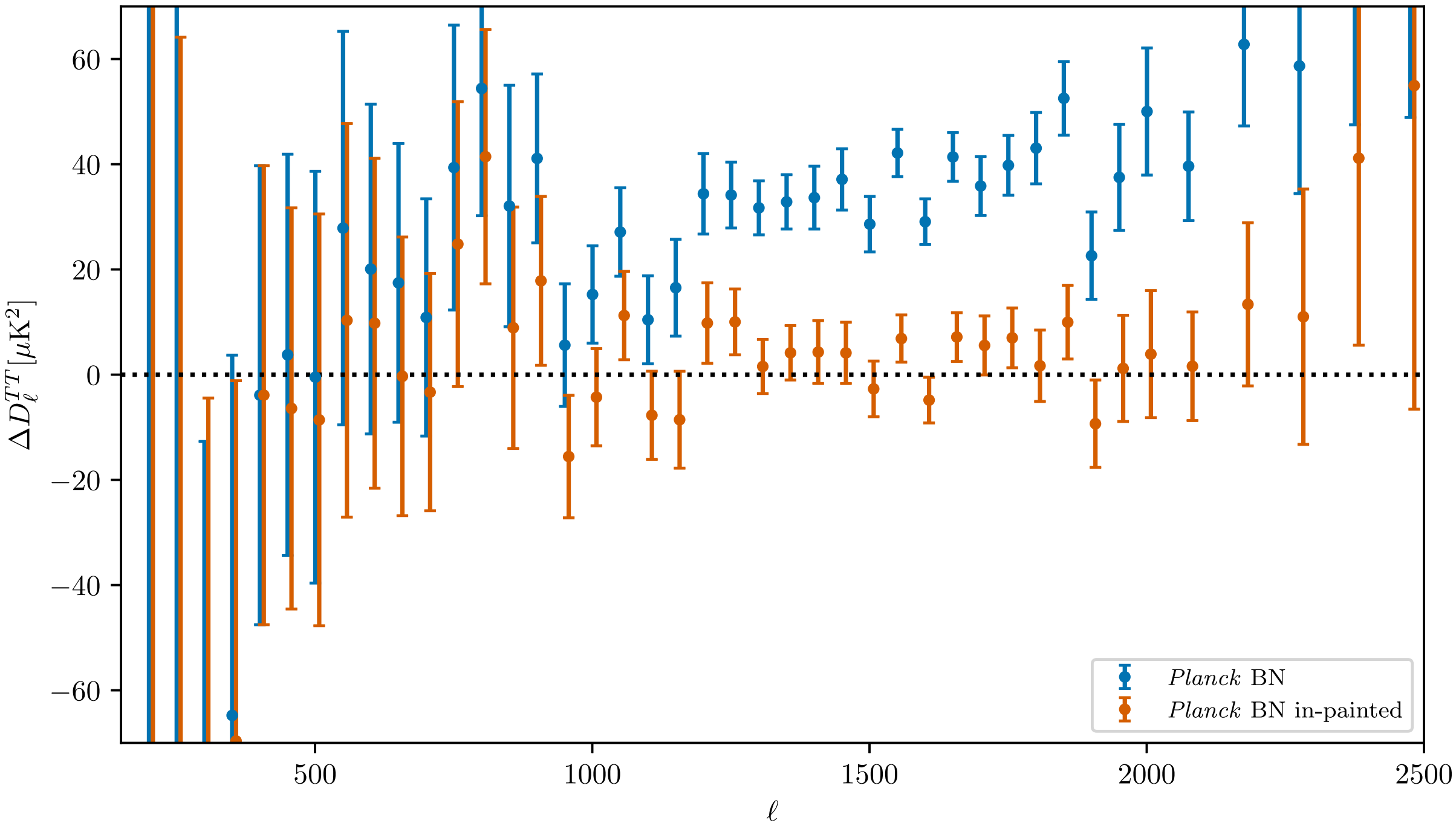

FIG. 10. As Fig. 5 for in-painted *Planck* maps on the BN patch. We plot the difference of *Planck* power spectra from filtered and unfiltered maps. The blue points are as in Fig. 5, while the orange points show spectra from maps that are in-painted before filtering.

A further complication we consider here is the case of unmasked sources. In ACT DR4 there are numerous sources that are projected out of the source-free maps but fall below the flux cut to be masked. When these sources are added back to the maps they will be unfiltered, and therefore boosted when the filter transfer function is divided out; this could potentially complicate the modeling of point sources in the likelihood. We again take the example of the BN patch to estimate the magnitude of this effect. For the autospectrum of the source-only map (again using s15/PA1), masked using the ACT BN footprint and ACT source mask, we find $C_\ell^{TT} = 1.4 \times 10^{-5}\ \mu\mathrm{K}^2$ from the unmasked sources. The boost in power from the filter transfer function is approximately linear in $D_\ell$, of modest amplitude: 0.23 $\mu\mathrm{K}^2$ at $\ell = 1000$ (cf. 2.23 $\mu\mathrm{K}^2$ before dividing out the transfer function) and 0.63 $\mu\mathrm{K}^2$ (cf. 20.05 $\mu\mathrm{K}^2$) at $\ell = 3000$. The amplitude may be somewhat higher (or lower) for other maps, but this small residual is also likely to be partially absorbed by the foreground parameters in a likelihood. Therefore, while this effect should be borne in mind, we do not expect it to be a serious impediment to the method.

## VI. CONCLUSIONS

We have studied in detail the interaction of Fourier-space filtering with point sources, focusing on the case of an ACT DR4-like analysis applied to ACT and *Planck* data. We find a significant bias, approximately linear in $D_\ell$, in *Planck* TT when filtered in this manner. This bias is due to cross-shaped artifacts coming from an interaction between masking and filtering, and scales with the point-source power in a particular patch; this can lead to sometimes surprising results, such as significant power spectrum differences between patches with differing point-source levels despite masking of those sources. The effect on polarization and *Planck* ×ACT spectra is much smaller due to lower point-source fluxes in polarization and the implementation of source subtraction for ACT. These artifacts also add some variance to the map even when there is no bias, but we find the magnitude to be small. Fortunately, in both temperature and polarization these effects can be relatively straightforwardly mitigated with multiple methods, including source subtraction before filtering and masking and in-painting of source holes. We rule these source-filter interactions out as the source of the ACT−*Planck* disagreement in TE raised in Choi *et al.* [18], but highlight them as important to account for in future analyses using *Planck* and other datasets where similar filtering is applied.

## ACKNOWLEDGMENTS

We are grateful to the ACT Collaboration for the data products used in this work and for helpful discussions and comments on the manuscript. Thanks in particular to Jo Dunkley, Thibaut Louis, Steve Choi, and the ACT power spectrum working group. E. R. acknowledges support from an Isaac Newton Studentship. A. C. acknowledges support from the STFC (Grants No. ST/W000997/1 and No. ST/X006387/1). This research used resources of the National Energy Research Scientific Computing Center, a U.S. Department of Energy Office of Science User Facility located at Lawrence Berkeley National Laboratory, operated under Contract No. DE-AC02-05CH11231. In this work we have used the following software packages: PSPY, PSPIPE, and PIXELL from [39], NAMASTER [40], CAMB [41,42], SCIPY [43], NUMPY [44], ASTROPY [45,46], MATPLOTLIB [47], HEALPY [48], HEALPIX [49], and ORPHICS [50].

## DATA AVAILABILITY

The data that support the findings of this article are openly available [51].